\documentclass[journal=nalefd,manuscript=letter,layout=twocolumn]{achemso}

\usepackage[version=3]{mhchem} 
\usepackage[T1]{fontenc}       
\usepackage[usenames]{color}
\usepackage[normalem]{ulem}
\usepackage{soul}
\usepackage[none]{hyphenat}

\newcommand{\corin}[1]{{\color{red} #1}}

\author{Toru Hirahara}
\email{hirahara@phys.titech.ac.jp}
\corin{\phone{+81 (0)3 5734 2365}}
\corin{\fax{+81 (0)3 5734 2365}}
\affiliation{Department of Physics, Tokyo Institute of Technology, Tokyo 152-8551, Japan}
\author{Sergey V. Eremeev}
 \affiliation{Institute of Strength Physics and Materials Science, Tomsk, 634055, Russia}
 \alsoaffiliation{Tomsk State University, Tomsk, 634050, Russia}
 \alsoaffiliation{Saint Petersburg State University, Saint Petersburg, 198504, Russia}
 \alsoaffiliation{Donostia International Physics Center (DIPC), Paseo de Manuel Lardizabal, 4, 20018 San Sebasti$\acute{a}$n/Donostia, Basque Country, Spain}
\author{Tetsuroh Shirasawa}
\affiliation{Institute for Solid State Physics, University of Tokyo, Kashiwa 277-8581, Japan}
\altaffiliation{Present address: National Institute of Advanced Industrial Science and Technology, Ibaraki 305-8560, Japan}
\author{Yuma Okuyama}
\affiliation{Department of Physics, Tokyo Institute of Technology, Tokyo 152-8551, Japan}
\author{Takayuki Kubo}
\affiliation{Department of Physics, University of Tokyo, Tokyo 113-0033, Japan}
\author{Ryosuke Nakanishi}
\affiliation{Department of Physics, University of Tokyo, Tokyo 113-0033, Japan}
\author{Ryota Akiyama}
\affiliation{Department of Physics, University of Tokyo, Tokyo 113-0033, Japan}
\author{Akari Takayama}
\affiliation{Department of Physics, University of Tokyo, Tokyo 113-0033, Japan}
\author{Tetsuya Hajiri}
\affiliation{UVSOR Facility, Institute for Molecular Science, Okazaki 444-8585, Japan}
\altaffiliation{Present address: Department of Crystalline Materials Science, Nagoya University, Nagoya 464-8603, Japan}
\author{Shin-ichiro Ideta}
\affiliation{UVSOR Facility, Institute for Molecular Science, Okazaki 444-8585, Japan}
\author{Masaharu Matsunami}
\affiliation{UVSOR Facility, Institute for Molecular Science, Okazaki 444-8585, Japan}
\altaffiliation{Present address: Energy Materials Laboratory, Toyota Technological Institute, Nagoya 468-8511, Japan}
\author{Kazuki Sumida}
\affiliation{Graduate School of Science, Hiroshima University, Higashi-Hiroshima 739-8526, Japan}
\author{Koji Miyamoto}
\affiliation{Hiroshima Synchrotron Radiation Center, Hiroshima University, Higashi-Hiroshima 739-8526, Japan}
\author{Yasumasa Takagi}
\affiliation{Department of Materials Molecular Science, Institute for Molecular Science, Okazaki 444-8585, Japan}
\author{Kiyohisa Tanaka}
\affiliation{UVSOR Facility, Institute for Molecular Science, Okazaki 444-8585, Japan}
\author{Taichi Okuda}
\affiliation{Hiroshima Synchrotron Radiation Center, Hiroshima University, Higashi-Hiroshima 739-8526, Japan}
\author{Toshihiko Yokoyama}
\affiliation{Department of Materials Molecular Science, Institute for Molecular Science, Okazaki 444-8585, Japan}
\author{Shin-ichi Kimura}
\affiliation{UVSOR Facility, Institute for Molecular Science, Okazaki 444-8585, Japan}
\altaffiliation{Present address: Graduate School of Frontier Biosciences and Department of Physics, Osaka University, Suita 565-0871, Japan}
\author{Shuji Hasegawa}
\affiliation{Department of Physics, University of Tokyo, Tokyo 113-0033, Japan}
\author{Evgueni V. Chulkov}
\affiliation{Donostia International Physics Center (DIPC), Paseo de Manuel Lardizabal, 4, 20018 San Sebasti$\acute{a}$n/Donostia, Basque Country, Spain}
\alsoaffiliation{Departamento de F$\acute{i}$sica de Materiales, Facultad de Ciencias Qu$\acute{i}$micas, UPV/EHU, Apdo. 1072, 20080 San Sebasti$\acute{a}$n, Basque Country, Spain}
\alsoaffiliation{Centro de F$\acute{i}$sica de Materiales, CFM-MPC, Centro Mixto CSIC-UPV/EHU, Apdo.1072, 20080 San Sebasti$\acute{a}$n/Donostia, Basque Country, Spain}
 \alsoaffiliation{Tomsk State University, Tomsk, 634050, Russia}
 \alsoaffiliation{Saint Petersburg State University, Saint Petersburg, 198504, Russia}
\title
  {A large-gap magnetic topological heterostructure formed by subsurface incorporation of a ferromagnetic layer}

\begin{document}

\begin{tocentry}

\includegraphics{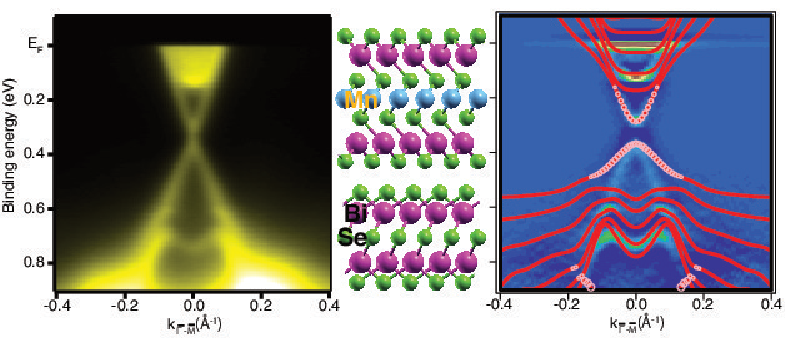}

\end{tocentry}

\begin{abstract}
 Inducing magnetism into topological insulators is intriguing for utilizing exotic phenomena such as the quantum anomalous Hall effect (QAHE) for technological applications. While most studies have focused on doping magnetic impurities to open a gap at the surface-state Dirac
point, many undesirable effects have been reported to appear in some cases
that makes it difficult to determine whether the gap opening is due
to the time-reversal symmetry breaking or not. Furthermore, the realization of the QAHE has been limited to low temperatures. Here we have
succeeded in generating a massive Dirac cone in a
MnBi$_2$Se$_4$/Bi$_2$Se$_3$ heterostructure which was fabricated by
self-assembling a MnBi$_2$Se$_4$ layer on top of the
Bi$_2$Se$_3$ surface as a result of the co-deposition of Mn and Se.
Our experimental results, supported by relativistic {\em ab initio}
calculations, demonstrate that the fabricated
MnBi$_2$Se$_4$/Bi$_2$Se$_3$ heterostructure shows ferromagnetism up
to room temperature and a clear Dirac-cone gap opening of
$\sim$100~meV without any other significant changes
in the rest of the band structure. It can be considered as a result
of the direct interaction of the surface Dirac cone and the magnetic
layer rather than a magnetic proximity effect. This spontaneously
formed self-assembled heterostructure with a massive Dirac spectrum,
characterized by a nontrivial Chern number 
$C=-1$, has a potential to realize the QAHE at
significantly higher temperatures than reported
up to now and can serve as a platform for developing future “topotronics” devices.\end{abstract}

\vspace{1cm}
Classification of materials and phases based on the ``topological
properties'' of the system has become one of the most extensively
studied research fields in physics and was the topic for the Nobel prize in physics in
2016 \cite{nobel2016}.
Topological insulators (TIs) are insulating
materials that have metallic surface states, whose electron spins are locked to their
momentum \cite{hasanreview,shenreview}. In the simplest case, the
surface states are helical Dirac fermions and the Dirac point is
robust due to the protection by time-reversal symmetry. When the
time-reversal symmetry is broken in TIs, novel quantum phenomena
have been predicted to occur including the formation of magnetic
monopoles \cite{monopole}, the quantum anomalous
Hall effect (QAHE) \cite{haldane}, and the topological magnetoelectric
effect \cite{topomag}.

In terms of the electronic structure,
a magnetic TI is expected to host a massive Dirac cone with a band gap.
Many researches have been performed to induce magnetism in TIs by magnetic impurity doping when growing
thin films \cite{hasan,xue2}. Although it
seemed that this was the most efficient way with the
successful observation of the QAHE in Cr or V-doped (Bi,Sb)$_2$Te$_3$ thin films
\cite{xue2,bestwick,chang}, the precise quantization of the Hall resistance at zero field is
still only limited to low temperature (10-100~mK). The temperature for the realization of the QAHE ($T_{QAHE}$) depends on the Curie temperature as well as the size of the Dirac cone gap of the system. Up to now, the main obstacle in enhancing $T_{QAHE}$ is an inhomogeneous distribution of magnetic atoms over the TI film that results in strong fluctuations of the magnetic energy gap \cite{davis,xuemat}. Modulation doping was shown to increase $T_{QAHE}$ and the QAHE was observed at 2~K for Cr-doped (Bi, Sb)$_2$Te$_3$ thin films when the magnetic-doped layer was placed 1~nm below the surface \cite{mogi}. A recent theoretical work suggests that this can be enhanced by V and I co-doping of Sb$_2$Te$_3$ \cite{qi}.

Another drawback of the magnetically doped systems is that there has
been no direct observation of the massive Dirac cone for the samples
that show the QAHE using angle-resolved photoemission spectroscopy
(ARPES) \cite{he}. With local probes such as scanning tunneling
spectroscopy (STS), an inhomogeneous Dirac-cone gap was observed for
Cr-doped (Bi, Sb)$_2$Te$_3$ \cite{davis}. On the other hand, for
V-doped Sb$_2$Te$_3$, it was reported that impurity states will
reside within the Dirac-cone gap and as a result, the density of
states will look gapless \cite{bode}. Moreover, impurity bands will
emerge near the Dirac point and it was reported that a
massive Dirac cone can arise even due to a nonmagnetic origin for
impurity-doped Bi$_2$Se$_3$ films with ARPES \cite{rader} . Thus one can say that
magnetic impurity doped TI films are not promising in terms of (i)
enhancing $T_{QAHE}$ to the region where application becomes
possible and (ii) clearly observing the gapped Dirac cone as a
result of the time-reversal symmetry breaking in a macroscopic
scale.

An alternative way to induce magnetism in TIs is to make a
heterostructure of TIs and magnets. Such systems are also used in
experiments to observe the surface-state spin transport \cite{shiomi,fert,samarth,wei}. But there has not been a direct
observation of the QAHE as well as the existence of a gapped Dirac
cone at the interfaces in these heterostructures. One reason for
this is the absence an atomically sharp interface that makes the gap
small and difficult to observe, which should be also responsible for
the low spin-charge conversion efficiency. Even in some cases with a
sharp interface (EuS/Bi$_2$Se$_3$), the Dirac-cone gap was shown not
to be related with the magnetic proximity effect \cite{sergeyeus},
although magnetization measurements showed evidence of a magnetic
interaction \cite{EuS}. In another case of MnSe/Bi$_2$Se$_3$, it was
shown from {\it ab initio} calculations that states other than the
gapped Dirac cone will cross $E_{\mathrm F}$ \cite{sergey} and make
the QAHE impossible to realize. However, the measured band structure
was completely different from the calculation but no clear
explanation was given \cite{Matetskiy}. Thus the
interplay between the topological properties and magnetism in an
ordered layered system needs to be examined further.

In the present study, we take advantage of self-organization to
produce an ideal system of a magnetic insulator/TI heterostructure,
which is formed spontaneously as a hexagonal MnBi$_2$Se$_4$ septuple
layer (SL) on the basis of the topmost quintuple layer (QL) of
Bi$_2$Se$_3$ film under Mn and Se co-deposition. Due to the
ferromagnetism of MnBi$_2$Se$_4$ up to room temperature as revealed
by superconducting quantum interference device (SQUID) and X-ray
magnetic circular dichroism (XMCD) measurements, the Dirac cone
becomes massive with a gap size of $\sim$ 100~meV. First-principles
calculations show that the calculated Chern number is $C=-1$ and the
heterostructure is identified as a quantum anomalous Hall phase. Our
results not only show the first example of a clear Dirac-cone gap
opening at the ferromagnet/TI interface, but imply that this should
be a suitable system to realize the QAHE at higher temperatures than
previously reported by tuning the Fermi level as well as to develop
novel devices based on the topological magnetoelectric effect.

\begin{figure*}[htbp]
\includegraphics[width=0.9\textwidth]{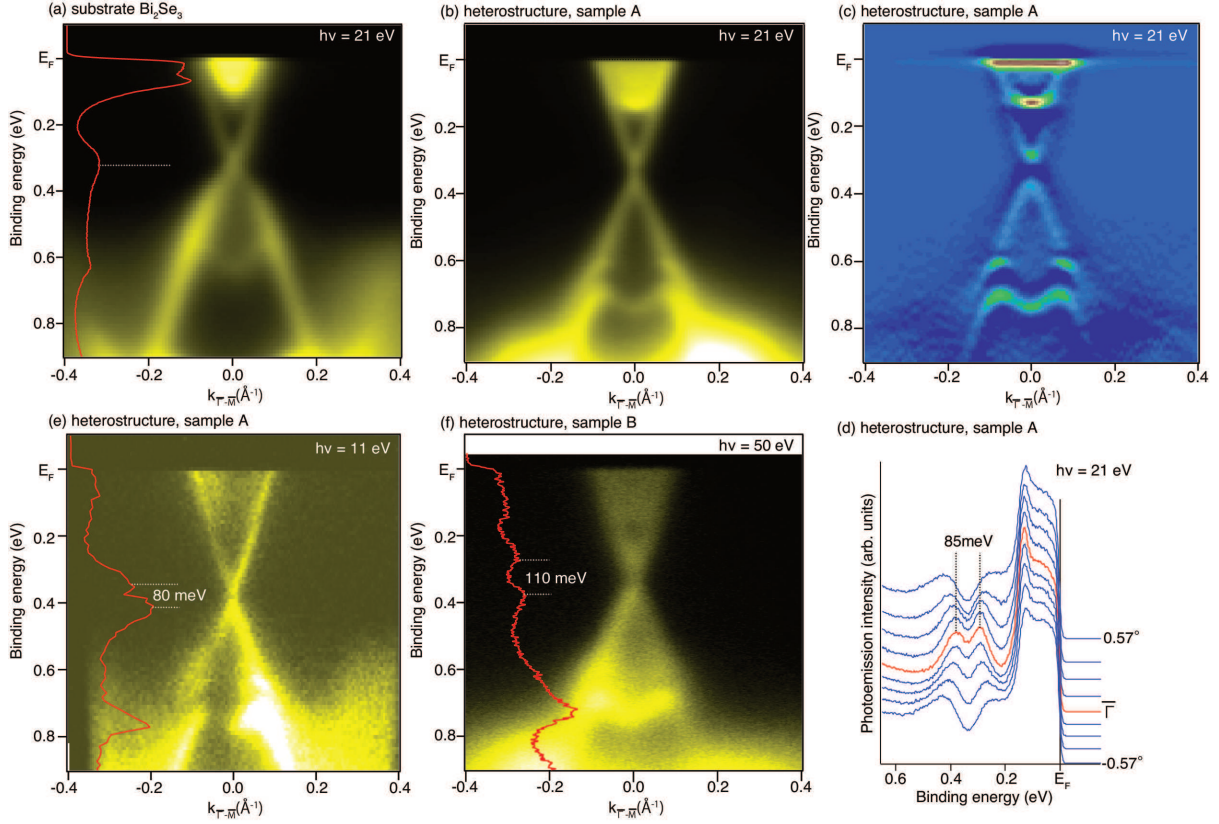}
\caption{\label{fig1} (a) Band dispersion of the
substrate Bi$_2$Se$_3$ film measured along the
$\bar{\mathrm{\Gamma}}-\bar{\mathrm{M}}$ direction taken at
$h\nu=21$~eV. The red line shows the
energy distribution curve (EDC) at the $\bar{\mathrm{\Gamma}}$ point. (b) Band dispersion of the heterostructure for sample
A measured along the $\bar{\mathrm{\Gamma}}-\bar{\mathrm{M}}$
direction taken at $h\nu=21$~eV. (c) The second derivative with
respect to the energy of the band dispersion in (b). (d) Raw spectra (EDCs)
near the $\bar{\mathrm{\Gamma}}$ point of the band dispersion in
(b). The gap size of the Dirac cone is 85~meV. (e) Band dispersion
of the heterostructure for sample A measured along the
$\bar{\mathrm{\Gamma}}-\bar{\mathrm{M}}$ direction taken at
$h\nu=11$~eV. The gap size of the Dirac cone is 80~meV. The red line shows the
EDC at the $\bar{\mathrm{\Gamma}}$ point. (f) Band
dispersion of the heterostructure for sample B measured along the
$\bar{\mathrm{\Gamma}}-\bar{\mathrm{M}}$ direction taken at
$h\nu=50$~eV. The gap size of the Dirac cone is 110~meV. The red line shows the
EDC at the $\bar{\mathrm{\Gamma}}$ point. All the measurements were performed at 30~K.}
\end{figure*}

Figure~\ref{fig1}(a) shows the band dispersion of the Bi$_2$Se$_3$
thin film which was used as the substrate for the heterostructure
formation. The well-known Dirac-cone of the surface states in
the bulk band gap can be seen together with the bulk conduction band
at the Fermi level due to the unintentional doping. The
Dirac point is recognized as shown in the energy distribution curve
(EDC) at the $\bar{\mathrm{\Gamma}}$ point (red line).
Figure~\ref{fig1}(b) shows the band structure of the heterostructure
(sample A) after Mn and Se were co-deposited on Bi$_2$Se$_3$. It was
measured immediately after cooling down the sample. A clear gap
opening is observed at the Dirac point. This becomes more prominent
by making the second derivative with respect to the energy of the
image in Fig.~\ref{fig1}(b), which is shown in Fig.~\ref{fig1}(c).
The gap size is deduced as $\sim$85~meV from the EDCs in
Fig.~\ref{fig1}(d). A slightly smaller gap is observed when the photon energy is
changed to 11~eV, as shown in Fig.~\ref{fig1}(e). The estimated gap
size is $\sim$80~meV, which is nearly the same as that shown in
Fig.~\ref{fig1}(d). It can also be noticed that the midpoint of the
energy gap has shifted down by about
70~meV from Figs.~\ref{fig1}(b)-(d) to Fig.~\ref{fig1}(e), due to
the band bending near the surface caused by the residual gas or
defect formation \cite{hofmann,hsieh}, since Fig.~\ref{fig1}(e) was
measured 12 hours after Figs.~\ref{fig1}(b)-(d). The band dispersion
images taken with other photon energies for sample A are shown in
Fig.~S1. Even though the gap exists
irrespective of the Fermi level position, the gap size slightly
changes depending on the measurement condition. However, there is no
correlation between the mid-gap position and the gap size.
Figure~\ref{fig1}(f) shows the band dispersion image of a different
sample (sample B) taken at $h\nu=50$~eV just after cooling down the
sample posterior to the preparation. The gap size is $\sim$110~meV,
larger than that for sample A although the midpoint of the gap is
nearly the same. We have also measured the gap size for other
samples and the gap value is in the range of 75-120~meV when the
Dirac point is located at the binding energy of 0.31-0.39~eV,
although there is no correlation between the two quantities. We
conclude that when Mn and Se are co-deposited on Bi$_2$Se$_3$ and a
heterostructure is formed, the Dirac cone becomes gapped. At first
glance, it should be due to the magnetic proximity effect from
antiferromagnetic MnSe.

However, there is one contradictory point about this band dispersion
of the heterostructure. According to {\it ab inito} calculations \cite{sergey}, the band
dispersion of a MnSe/Bi$_2$Se$_3$ heterostructure should have the
following characteristics: (i) a massive Dirac cone with a gap of
8.5~meV, (ii) other metallic states in the Bi$_2$Se$_3$ bulk band
gap. In our experimental data, we do not find any additional
states in the Dirac state energy region. Moreover, the gap size
is an order of magnitude larger than that predicted in the
calculation. Therefore, the calculation and the experiment are not
consistent with each other. Various factors may be responsible for
this inconsistency, such as the difference in the interface
structure or the thickness of the MnSe layers. We have therefore
performed structure analyses of the heterostructure based on the
LEED {\it IV} technique to resolve this inconsistency.

\begin{figure*}[htbp]
\includegraphics[width=0.85\textwidth]{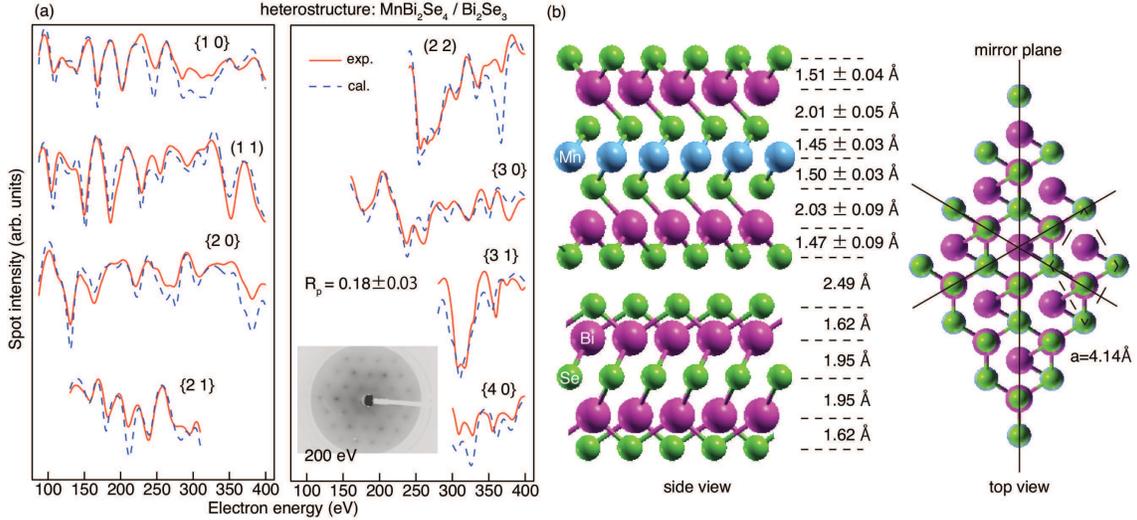}
\caption{\label{fig2} (a) Experimental {\it IV}
spectra of LEED spots measured at 100~K for the heterostructure (red solid lines), and the
calculated spectra of the optimized model shown in (b) (blue dotted lines). The inset
shows the LEED pattern at 200~eV. (b) Cross-sectional (left) and top (right) view of the
optimized model of the heterostructure, which turns out to be
MnBi$_2$Se$_4$/Bi$_2$Se$_3$. Blue, pink, and green atoms represent
Mn, Bi, and Se respectively. The solid lines show the mirror plane and the
dotted parallelogram shows the unit cell.}
\end{figure*}

Figure~\ref{fig2}(a) shows the experimentally observed {\it IV}
curves (red solid lines). To reproduce these curves, we have first assumed the
MnSe/Bi$_2$Se$_3$ heterostructure that is expected from the sample
preparation method. However, we could not obtain R-factor ($R_p$)
values smaller than 0.4. Thus it was proven that the heterostructure
we have fabricated was not MnSe/Bi$_2$Se$_3$. We tried other
structures and finally found one that was in striking agreement with
the experimental LEED {\it IV} curves. The calculated {\it IV}
curves for the optimized structure are shown by the blue-dotted
lines in Fig.~\ref{fig2}(a). The agreement between the two curves is
excellent with $R_p=0.18\pm0.03$. The structure thus determined is
shown in Fig.~\ref{fig2}(b). What is remarkable about this structure
is that the deposited Mn and Se are not on top of the
Bi$_2$Se$_3$ substrate, but they are
incorporated inside the topmost QL of Bi$_2$Se$_3$. So we can say
that a MnBi$_2$Se$_4$ septuple layer (SL) spontaneously forms on top
of Bi$_2$Se$_3$ by self-organization. Although the microscopic
mechanism of MnSe bilayer immersion into the QL is not clear, our
calculations show that the MnBi$_2$Se$_4$ SL is 630 meV
energetically more favorable as compared to the MnSe bilayer
(BL)/Bi$_2$Se$_3$-QL interface. MnBi$_2$Se$_4$ in the bulk form is
reported to have the monoclinic structure with {\it C2/m} symmetry.
It is shown to be a narrow gap semiconductor and antiferromagnetic
with $T_{Neel}\sim14$~K \cite{MBS}. However, the MnBi$_2$Se$_4$ SL
that we have fabricated has a hexagonal structure and is a
semiconductor with a gap of $\sim0.4$~eV (Fig.~S3).

\begin{figure*}[htbp]
\includegraphics[width=0.9\textwidth]{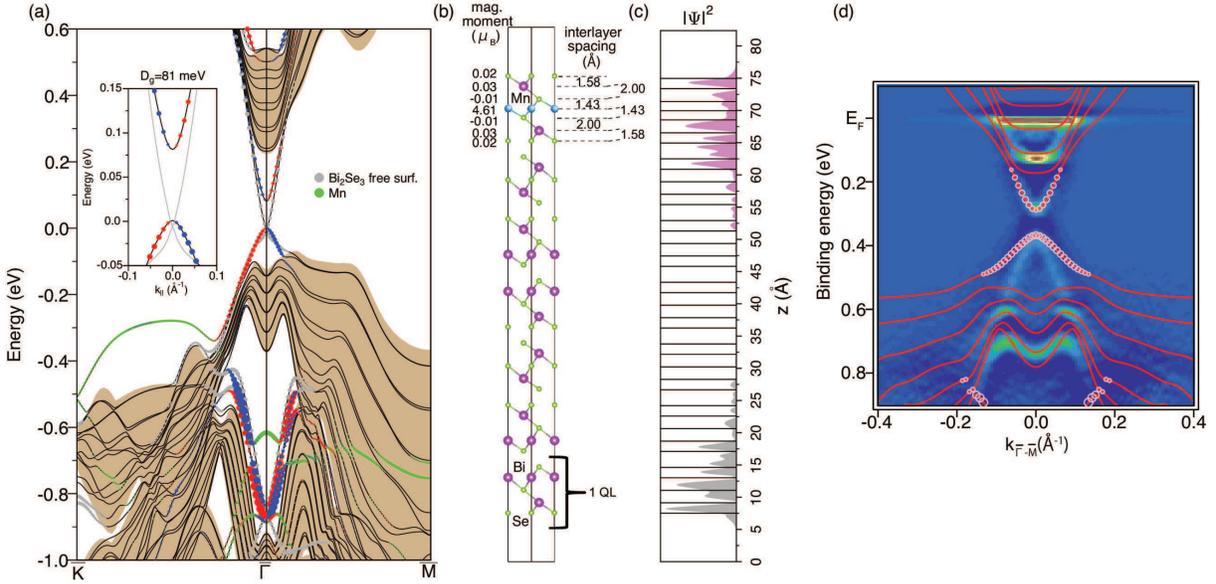}
\caption{\label{fig3} (a) Calculated band
dispersion of a MnBi$_2$Se$_4$/6~QL Bi$_2$Se$_3$ heterostructure.
The states marked in red and blue are localized at the topmost SL
and the next Bi$_2$Se$_3$ QL and spin-polarized in the in-plane
direction perpendicular to the wavenumber (positive and negative,
respectively). The states marked in green are the ones mostly
localized at MnBi$_2$Se$_4$ SL and labeled ``Mn''. The states
localized at the bottom surface of the slab are indicated in gray.
The shaded area shows the bulk band projection of Bi$_2$Se$_3$. The
inset shows a close-up of the Dirac-state dispersion. A gap of 81~meV is
found for the Dirac cone at the top surface. (b) The real space
atomic configuration used in the calculation together with the
interlayer spacings and layer-resolved magnetic moments. (c)
Wave-function ($|\Psi|^2$) plot for the
gapped Dirac state at the top surface
(violet), and that at the bottom surface
(gray), respectively. (d) The calculated band dispersion
overlapped with the experimental data of Fig.~\ref{fig1} (c).
Pink circles show the states localized in the SL.}
\end{figure*}

The electronic structure of the heterostructure was calculated based
on the determined atomic structure of Fig.~\ref{fig2}(b). Namely, a
six QL film of Bi$_2$Se$_3$
was sandwiched between MnBi$_2$Se$_4$ and vacuum. The interlayer
spacing values obtained after structural relaxation of
the heterostructure and shown in
Fig.~\ref{fig3}(b) are in excellent agreement with the
experimentally determined values shown in Fig.~\ref{fig2}(b). We
found that the spin structure with the Mn magnetic moment pointing
out-of-plane is 0.2~meV energetically more favorable than the one
with in-plane spin polarization and the magnetic moment of Mn was
found to be 4.61~$\mu_B$ (Fig.~\ref{fig3}(b)). The calculated band
dispersion is shown in Fig.~\ref{fig3}(a). The red and blue points
correspond to the states with in-plane spin polarization
perpendicular to the wavenumber (positive and negative,
respectively) that are localized at the topmost SL and the
neighboring QL (refer to the supplementary material for the
experimental results of the Dirac-cone spin-polarization). Near the
Fermi level, the massless Dirac cone of pristine Bi$_2$Se$_3$ at the
bottom surface can be found (gray points and gray line in inset)
together with the massive Dirac state with a gap D$_g$=81~meV at the
top surface. The ``gray'' band can be a good reference for tracking
the changes in the Dirac state. As can be seen, the lower part of
the Dirac cone of the gapped state shows almost no change in the
energy position and the gap is formed by shifting up the upper
branch. The calculations performed for a symmetric heterostructure,
where Bi$_2$Se$_3$ is sandwiched between two MnBi$_2$Se$_4$ SLs,
result in the same value of the Dirac-cone gap. Figure~\ref{fig3}(d)
shows this calculated spectrum overlapped with the experimental
result (the calculation was rigidly shifted by -0.38~eV to fit to
the experimental $E_{\mathrm F}$). The agreement between the two is excellent with exception of flatter dispersion of the bottom part of the Dirac cone in the calculation whereas the gap size in the Dirac cone, as well as fine structures other than the Dirac cone, are well reproduced. The green points in Fig.~\ref{fig3}(a) are states localized at the topmost SL
and labeled ``Mn'' because it has the maximum contribution from the
Mn $p_z$ orbitals. They are located along the
$\bar{\mathrm{\Gamma}}-\bar{\mathrm{K}}$ direction in the
Bi$_2$Se$_3$ bulk band gap and a band that resembles this is also
found in the calculated dispersion of the free-standing
MnBi$_2$Se$_4$ SL (Fig.~S3). This was actually observed
experimentally when we performed ARPES measurements along the
$\bar{\mathrm{\Gamma}}-\bar{\mathrm{K}}$, as shown in Fig.~S9(a).
Altogether, the calculation reproduces the experimentally observed
band structure nicely, and we are sure that the heterostructure we
have fabricated is MnBi$_2$Se$_4$/Bi$_2$Se$_3$. Moreover, one can
say that the mass acquisition of the Dirac cone is not due to a
magnetic proximity effect because the Dirac state and the magnetic
layer are both at the topmost SL (70\% of the Dirac state is
localized in the SL, see Fig.~\ref{fig3}(c)). We should rather say
that this is a ``direct interaction'' between the magnetic layer and
the surface Dirac cone, since they overlap in real space and some part of the wave-function of the Dirac state is within the Mn layers. The original Dirac cone of Bi$_2$Se$_3$ has been pushed up to the newly formed MnBi$_2$Se$_4$ capping layer, similar to that in other topological-insulator based heterostructures discussed in Ref.~29.
This is the reason why there is a large gap
of $\sim$80~meV at the Dirac point, in contrast to the case of
MnSe/Bi$_2$Se$_3$ where a magnetic proximity effect was considered
\cite{sergey}.

The gapped Dirac state has a potential to realize the QAHE. The
quantized Hall conductance $\sigma_{yx}=Ce^2/h$ is related to the topological
characteristic of the band structure known as the Chern number $C$
\cite{Thouless}. Our calculations show that the
gapped spectrum of the MnBi$_2$Se$_4$/Bi$_2$Se$_3$ heterostructure
is characterized by the Chern number $C=-1$, that defines the
obtained heterostructure unambiguously as a quantum
anomalous Hall phase. To confirm the topological
character of the heterostructure spectrum, we artificially decreased
the spin-orbit interaction strength $\lambda$ and found that it
leads to the gap narrowing and its closing at $\lambda/\lambda_0
\approx 0.75$ (see Fig.~S4). Upon further decrease of the
spin-orbit interaction strength, the gap reopens and the system
becomes a topologically trivial insulator.

\begin{figure*}[htbp]
\includegraphics[width=0.9\textwidth]{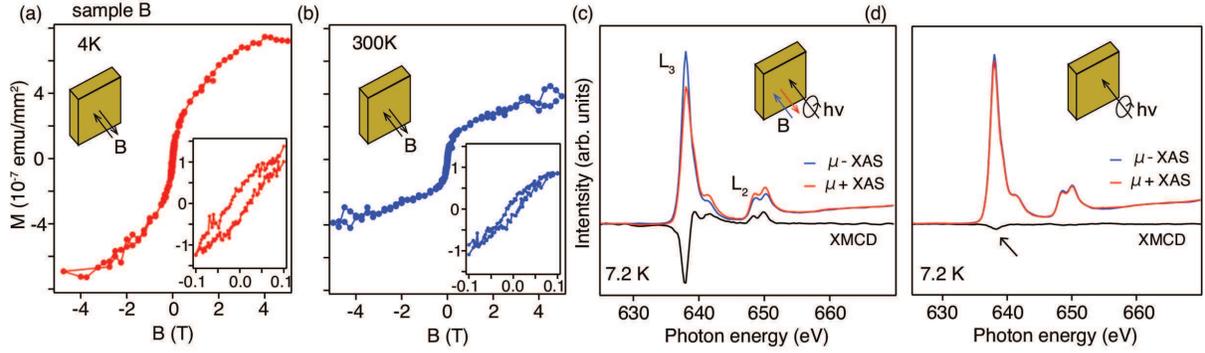}
\caption{\label{fig4} (a),(b) Magnetization curves
measured with SQUID for the Se-capped sample B taken at 4~K (a) and at 300~K (b), respectively. The magnetic field was applied
perpendicular to the sample surface. The inset shows the close-up near zero-field. Clear hysteresis loops can be observed both at 4~K and
300~K. (c) X-ray absorption spectra (XAS) at 7.2~K for a circularly
polarized incident light when a $\pm$5~T magnetic field was applied
along the surface-normal direction. $\mu_+$ and $\mu_-$ correspond to the spectrum obtained at
+5~T and $-$5~T, respectively. The corresponding XMCD spectra
is also shown. The Se capping layer was removed by annealing the
sample in UHV. (d) Same as (c) but measured at zero field. A remanent XMCD
signal can be observed (shown by the arrow). }
\end{figure*}

Next, we have tried to experimentally verify the absence of
time-reversal symmetry in the heterostructure as the calculation
suggests. Figures~\ref{fig4}(a) and (b) show the magnetization
curves measured with SQUID for the Se-capped sample B taken at 4~K
(a) and at 300~K (b), respectively. The magnetic field was applied
perpendicular to the sample surface (The data for the in-plane magnetization can be found in Fig.~S7). In order to
display the situation near zero-field clearly, we show in the inset the
close-up for -0.1 to 0.1~T. There is a vivid hysteresis loop both at
4~K and 300~K, demonstrating the ferromagnetic nature of the
heterostructure.\cite{note} Actually, the gapped Dirac cone can also be observed at room temperature (Fig.~S2) and thus the Curie temperature is confirmed to be above 300~K also from ARPES.
We also performed XMCD measurements after removing the Se cap in UHV to make sure that the Se cap is not affecting the magnetic properties.
Figure~\ref{fig4}(c) shows the X-ray absorption (XAS) spectra at the Mn $L$
edge measured at 7.2~K for a circularly polarized incident light
when a $\pm$5~T magnetic field was applied along the surface-normal
direction. $\mu_+$ and $\mu_-$ correspond to the spectrum obtained at
+5~T and -5~T, respectively.
The corresponding XMCD spectrum is also shown and there is
a clear signal. The quantitative analysis of the XMCD measurements using the sum-rule is discussed in the supplementary material, see Fig.~S5.
The measurements were also performed at remanence (0~T applied
magnetic field), namely the signal was obtained after switching off
the $\pm$5~T magnetic field applied perpendicular to the surface
(Fig.~\ref{fig4}(d)). Although weak (the peak at the $L_3$ edge is
$\sim$7\% compared to the data of Fig.~\ref{fig4}(c)), we still can
notice the XMCD signal (the arrow in Fig.~\ref{fig4}(d)) and this
confirms that the system is ferromagnetic. We
believe that the SQUID data of Figs.~\ref{fig4}(a) and (b) and that
obtained by XMCD in Figs.~\ref{fig4}(c) and (d) are both showing the
ferromagnetism of the Mn layer despite the different probing depth
of the two measurements, since it can be the only source of
ferromagnetism in the present heterostructure system.

The heterostructure we have fabricated demonstrates various advantages
compared to previously studied quantum anomalous Hall systems. First, it forms spontaneously by
depositing Mn and Se on top of Bi$_2$Se$_3$. We have done this using thin films, but in
principle it should also be possible to do this for single crystals or even thinner films. This is in contrast to the
extensively studied systems with an intentional magnetic impurity
doping \cite{hasan,rader,bode,xue2,bestwick}, in which fine-tuning
of the impurity concentration is needed to achieve a long range
magnetic order \cite{xue3}.

Secondly, in our samples, we
have an ideal situation where the Dirac cone and the magnetic layer
interact strongly due to the spatial overlap of the respective wave
functions, thus leading to a large magnetic gap but with minimal
effects on other bands. This is in contrast to heterostructures that utilize
the magnetic proximity effect \cite{sergey,sergeyeus,EuS}.

Thirdly, our samples are free from inhomogeneity since the structure was
clearly determined by LEED measurements. This allowed us to observe a clear Dirac cone gap with
ARPES and assign its origin unambiguously as due to the time-reversal symmetry
breaking. Recalling that the presence of the impurity states makes
it difficult to determine the gap opening in the Dirac cone as well
as its origin \cite{rader,bode}, this is a merit. Furthermore, the
clear gap opening in ARPES means that our heterostructure does not
show strong fluctuations of the magnetic energy gap \cite{davis}.

Finally, the Curie temperature of the heterostructure is above room
temperature. Furthermore, the Dirac-cone gap is larger than the
thermal fluctuation at room temperature. Therefore, we can expect to
observe the QAHE for our samples at higher temperatures than those
previously reported \cite{xue2,bestwick,chang} by moving the Fermi
level into the induced gap either chemically \cite{hasan} or by
gating \cite{xue2,chang,xue3}. 
So our heterostructure may be used in application for developing ``topotronics'' devices.

{\it Experimental and calculation methods}

The heterostructure samples were prepared by molecular beam epitaxy
in ultrahigh vacuum (UHV) chambers equipped with a
reflection-high-energy electron diffraction (RHEED) system. First,
a clean Si(111)-$7\times7$ surface was prepared on an $n$-type
substrate by a cycle of resistive heat treatments. The
Si(111)$\beta\sqrt{3}\times\sqrt{3}$-Bi surface was formed by 1~ML
(7.83$\times10^{14}$~cm$^{-2}$) of Bi deposition on the $7\times7$
surface at 620~K monitored by RHEED. Then Bi was deposited on the
$\beta\sqrt{3}\times\sqrt{3}$-Bi structure at $\sim$200~$^\circ$C in
a Se-rich condition. Such a procedure is reported to result in a
smooth epitaxial film formation with the stoichiometric ratio of
Bi:Se=2:3 \cite{zhang,sakamoto}. The grown Bi$_2$Se$_3$ films were
annealed at $\sim$240~$^\circ$C for 5 minutes. The thickness of the
Bi$_2$Se$_3$ films in this work is $\sim$ 15~QL, which does not show
the gap opening in the Dirac cone due to finite-size effects
\cite{sakamoto,xue}, as shown in Fig.~\ref{fig1}(a). Finally, Mn was
deposited on Bi$_2$Se$_3$ in a Se-rich condition at
$\sim$240~$^\circ$C. As reported in Ref.~\cite{Matetskiy}, this
results in a flat film formation which was believed to be a
heterostructure of MnSe/Bi$_2$Se$_3$.

ARPES and SARPES measurements were performed {\it in situ} after the
sample preparation. The ARPES measurements were performed at BL-5U
and 7U of UVSOR-III using $p$-polarized photons in the energy range
of 30-65~eV and 7.5-28~eV, respectively \cite{samrai}. All the data
shown were taken at 30~K unless otherwise indicated. The energy and
angular resolutions were 15~meV and 0.25$^\circ$, respectively. The
SARPES measurements were performed at BL-9B of HiSOR using
$p$-polarized photons in the energy of 18-21~eV at
30~K \cite{okuda}. In some of the measurements, a magnetic field as large as
$\sim$0.1~T was applied to the sample.
The energy and angular resolutions were 30~meV and 0.75$^\circ$, respectively.

LEED measurements were also performed {\it in situ} after the sample
formation in another UHV chamber. The LEED patterns with incident energy from 80 to 400 eV
were recorded in steps of 1 eV by a digital CCD camera at 100~K.
{\it IV} curves of 14 inequivalent diffraction spots were obtained.
In order to determine the surface structure, we calculated the {\it
IV} curves in the tensor LEED to fit the experimental {\it IV}
curves using the SATLEED package of Barbieri/Van Hove \cite{vanhove}
and minimized Pendry's R-factor ($R_p$). As shown in
Fig.~\ref{fig2}(b), each atomic layer was treated differently
according to their environments. The in-plane lattice constant of
the heterostructure was determined from positions of the LEED spots
as well as the RHEED spots and was the same with that of
Bi$_2$Se$_3$. Angular momentum up to 17 ($l_{max}$ = 17) was taken
into account because of the strong scattering of the heavy Bi atom
($Z= 83$). Considering the mean penetration depth of the incident
electrons of $\sim10~\mathrm{\AA}$, only the topmost 7 surface
layers were allowed to relax and we used the bulk Bi$_2$Se$_3$
parameters for the layers beneath. In search of the optimal
structure, the Debye temperature of each atom was changed in steps
of 10 K from 50 K up to 300 K. If MnSe grew on Bi$_2$Se$_3$, it
should have the {\it p3m1} symmetry. However, the symmetrically
inequivalent spots, such as (1,0) and (0,1) spots, exhibited almost
the same {\it IV} curves. Since this was also seen for the pristine
Bi$_2$Se$_3$ LEED patterns, it probably comes from the fact that
there are twin domains on the surface which are related by a
180$^\circ$ rotation. The superposition of the two domains should
lead to the apparent twofold symmetry. Taking this double-domain
surface into account, we took the average of the {\it IV} curves
both in the calculation and in the experimental data such that {\it
\{h,k\}} is the average of ($h$,$k$) and ($k$,$h$) spots (see
Fig.~\ref{fig2}(a)). Note that spots having the same mirror indices
of {\it h} and {\it k} do not need averaging.

For the SQUID measurements, the fabricated samples were capped with
$\sim$15~nm of Se before taking it out of the UHV chamber
and a commercial MPMS-52 system (Quantum Design)
was used. The diamagnetic contribution of the Si substrate was
derived from the field-dependent magnetization curves at room
temperature and subtracted from all data.

XMCD measurements were
performed at BL-4B of UVSOR-III with circularly polarized X-ray
radiation (positive helicity with the polarization of 0.6) at 7.2~K and a magnetic field as high as $\pm$5~T was applied
to the sample \cite{nakagawa}.  Instead
of changing the photon polarization, we reversed the direction of
the magnetic field to derive the XMCD spectra.
As shown in Fig.~\ref{fig4}(c), +(-) magnetic field
corresponds to the direction antiparallel (parallel)
to the direction of the photon incidence.
The measurements were first performed for
the samples with Se-capping. After measuring, the
Se-capped samples were annealed at $\sim$240~$^\circ$C to remove the
capping layers. LEED observations showed a clear recovery of the
$1\times1$ surface periodicity for the cap-removed samples.

For structural optimization and electronic band calculations we used
the Vienna Ab Initio Simulation Package \cite{kresse1,kresse2} with
generalized gradient approximation (GGA-PBE) \cite{perdew} to the
exchange-correlation potential. The interaction between the ion
cores and valence electrons was described by the projector
augmented-wave method \cite{blochl,kresse3}. The Hamiltonian
contains scalar relativistic corrections, and the spin-orbit
interaction (SOI) is taken into account by the second variation
method \cite{koelling}. To correctly describe the highly correlated
Mn-$d$ electrons we include the correlation effects within the
GGA+$U$ method \cite{liechtenstein}. The
in-plane lattice constant of the heterostructure was fixed to that
of Bi$_2$Se$_3$. DFT-D3 van der Walls corrections \cite{Grimme} were
applied for accurate structure optimization. The Chern number
was calculated for a symmetric
MnBi$_2$Se$_4$/5QL-Bi$_2$Se$_3$/MnBi$_2$Se$_4$ slab by using the
method based on tracking the evolution of hybrid Wannier functions
realized in Z2Pack \cite{Soluyanov}.

\begin{acknowledgement}
The authors thank A. V. Matetskiy, A. A. Saranin, O.
Rader, and A. Kimura for discussions. This work has been supported by Grants-In-Aid
from Japan Society for the Promotion of Science (Nos. 15H05453,
16K13683, 19340078, 23244066), the Toray Science Foundation, the
Basque Country Government, Departamento de Educaci\'{o}n,
Universidades e Investigaci\'{o}n (Grant No. IT-756-13), the Spanish
Ministry of Science and Innovation (Grant Nos. FIS2010-19609-C02-01,
FIS2013-48286-C02-02-P, and FIS2013-48286-C02-01-P), the Tomsk State
University Academic D.I. Mendeleev Fund Program (Grant No.
8.1.05.2015), and Saint Petersburg State University (project
15.61.202.2015). The ARPES experiments were performed under the
UVSOR Proposal Nos. 25-808, 26-531, 27-533, 28-526, and S-15-MS-0034
and the SARPES experiments were performed under the HiSOR Proposal
No. 15-A-14. The XMCD measurements were performed under the UVSOR
proposal number S-16-MS-2017. The LEED measurements were performed
under the ISSP Proposal number H17-A250. The SQUID measurements were
performed using facilities of the Cryogenic Research Center, the
University of Tokyo. Calculations were performed on the SKIF-Cyberia
supercomputer at the National Research Tomsk State University.

\end{acknowledgement}

\begin{suppinfo}
Additional details on ARPES, SARPES, {\it ab inito} calculations, SQUID, and XMCD measurements.
This material is available free of charge via the Internet at http://pubs.acs.org.
\end{suppinfo}


\begin{thebibliography}{99}
{
\bibitem{nobel2016}\begin{verb}
https://www.nobelprize.org/nobel_prizes/physics/laureates/2016/
\end{verb}
 \bibitem{hasanreview} Hasan, M. Z.; Kane, C. L., Colloquium: Topological insulators. {\it Rev. Mod. Phys.} {\bf 2010} {\it 82}, 3045.
 \bibitem{shenreview} Qi, X.-L.; Zhang, S.-C. Topological insulators and superconductors. {\it Rev. Mod. Phys.} {\bf 2011} {\it 83}, 1057.
 \bibitem{monopole} Qi, X.-L.; Li, R.; Zang, J.; Zhang, S.-C., Inducing a Magnetic Monopole with Topological Surface States. {\it Science} {\bf 2009} {\it 323} 1184.
 \bibitem{haldane} Haldane; F. D. M., Model for a quantum Hall effect without Landau levels: Condensed-matter realization of the parity anomaly. {\it Phys. Rev. Lett.} {\bf 1988} {\it 61}, 2015.
 \bibitem{topomag} Qi, X.-L.; Hughes, T. L.; Zhang, S.-C; Topological field theory of time-reversal invariant insulators. {\it Phys. Rev. B} {\bf 2008} {\it 78}, 195424.
 \bibitem{hasan} Xu, S.-Y. et al. Hedgehog spin texture and Berry's phase tuning in a magnetic topological insulator. {\it Nature Phys.} {\bf 2012} {\it 8}, 616.
 \bibitem{xue2} Chang, C.-Z. et al. Experimental Observation of the Quantum Anomalous Hall Effect in a Magnetic Topological Insulator. {\it Science} {\bf 2013} {\it 340}, 167.
 \bibitem{bestwick} Bestwick, A. J. et al. Precise Quantization of the Anomalous Hall Effect near Zero Magnetic Field. {\it Phys. Rev. Lett.} {\bf 2015} {\it 114}, 187201.
 \bibitem{chang} Chang C.-Z. et al. High-precision realization of robust quantum anomalous Hall state in a hard ferromagnetic topological insulator. {\it Nat. Mat.} {\bf 2015} {\it 14}, 473.
 \bibitem{davis} Lee, I. et al. Imaging Dirac-mass disorder from magnetic dopant atoms in the ferromagnetic topological insulator Cr$_x$(Bi$_{0.1}$Sb$_{0.9}$)$_{2-x}$Te$_3$. {\it Proceeding of the National Academy of Sciences} {\bf 2015} {\it 112}, 1316.
 \bibitem{xuemat} Feng, X. et al. Thickness Dependence of the Quantum Anomalous Hall Effect in Magnetic Topological Insulator Films. {\it Adv. Mater.} {\bf 2016} {\it 28}, 6386.
 \bibitem{mogi} Mogi, M. et al. Magnetic modulation doping in topological insulators toward higher-temperature quantum anomalous Hall effect. {\it Appl. Phys. Lett.} {\bf 2015} {\it 107}, 182401.
 \bibitem{qi} Qi, S. et al. High-Temperature Quantum Anomalous Hall Effect in $n$-$p$ Codoped Topological Insulators. {\it Phys. Rev. Lett.} {\bf 2016} {\it 117}, 056804.
 \bibitem{he} Chang C.-Z. et al. Thin Films of Magnetically Doped Topological Insulator with Carrier-Independent Long-Range Ferromagnetic Order. {\it Adv. Mater.} {\it 2013} {\bf 25}, 1065.
 \bibitem{bode} Sessi, P. et al. Dual nature of magnetic dopants and competing trends in topological insulators. {\it Nature Commun.} {\bf 2016} {\it 7}, 12027.
 \bibitem{rader} S\`anchez-Barriga J. et al. Nonmagnetic band gap at the Dirac point of the magnetic topological insulator (Bi$_{1-x}$Mn$_x$)$_2$Se$_3$. {\it Nature Commun.} {\bf 2016} {\it 7}, 10559.
 \bibitem{shiomi} Shiomi, Y. et al. Spin-Electricity Conversion Induced by Spin Injection into Topological Insulators. {\it Phys. Rev. Lett.} {\bf 2014} {\it 113}, 196601.
 \bibitem{fert} Rojas-S\'anchez, J.-C. et al. Spin to Charge Conversion at Room Temperature by Spin Pumping into a New Type of Topological Insulator: $\alpha$-Sn Films. {\it Phys. Rev. Lett.} {\bf 2016}, {\it 116}, 096602.
 \bibitem{samarth} Wang, H. et al. Surface-State-Dominated Spin-Charge Current Conversion in Topological-Insulator-Ferromagnetic-Insulator Heterostructures. {\it Phys. Rev. Lett.} {\bf 2016} {\it 117}, 076601.
 \bibitem{wei} Wei, P.; Katmis, F.; Assaf, B. A.; Steinberg, H.; Jarillo-Herrero, P.; Heiman, D.; and Moodera, J. S.; {\it Phys. Rev. Lett.} {\bf 2013} {\it 110}, 186807.
 \bibitem{sergeyeus} Eremeev, S. V.; Men'shov, V. N.; Tugushev, V. V.; Chulkov, E. V., Interface induced states at the boundary between a 3D topological insulator Bi$_2$Se$_3$ and a ferromagnetic insulator EuS. {\it J. Magn. Magn. Mater.} {\bf 2015} {\it 383}, 30.
 \bibitem{EuS} Katmis, F. et al. A high-temperature ferromagnetic topological insulating phase by proximity coupling. {\it Nature} {\bf 2016} {\it 533}, 513.
 \bibitem{sergey} Eremeev, S. V., Men'shov, V. N., Tugushev, V. V., Echenique, P. M. \& Chulkov, E. V. Magnetic proximity effect at the three-dimensional topological insulator/magnetic insulator interface. {\it Phys. Rev. B} {\bf 2013} {\it 88}, 144430.
 \bibitem{Matetskiy} Matetskiy, A. V. et al. Direct observation of a gap opening in topological interface states of MnSe/Bi$_2$Se$_3$ heterostructure. {\it Appl. Phys. Lett.} {\bf 2015} {\it 107}, 091604.
 \bibitem{hsieh} Hsieh, D. et al. A tunable topological insulator in the spin helical Dirac transport regime. {\it Nature} {\bf 2009} {\it 460}, 1101.
 \bibitem{hofmann} Bianchi, M. et al. Coexistence of the topological state and a two-dimensional electron gas on the surface of Bi$_2$Se$_3$. {\it Nature Commun.} {\bf 2010} {\it 1}, 128.
 \bibitem{MBS} Nowka, C. et al. Chemical vapor transport and characterization of MnBi$_2$Se$_4$. {\it J. Cryst. Growth} {\bf 2017} {\it 459}, 81.
\bibitem{hetero} Menshchikova, T. V., Otrokov, M. M., Tsirkin, S. S., Samorokov, D. A., Bebneva, V. V.,  Ernst, A., Kuznetsov, V. M., Chulkov, E. V., {\it Nano Lett.} {\bf 2013} {\it 13} 6064.
 \bibitem{Thouless} Thouless, D. J.; Kohmoto, M.; Nightingale, M. P.; den Nijs, M., Quantized Hall Conductance in a Two-Dimensional Periodic Potential, {\it Phys. Rev. Lett.} {\bf 1982} {\it 49}, 405.
\bibitem{note} The magnetization curve does not seem to saturating even at 5~T and shows a characteristic somewhat similar to a paramagnetic one. One possible explanation for this is the presence of Mn atoms at substitutional sites that give a paramagnetic contribution. 
 \bibitem{xue3} Chang, C.-Z. et al. Chemical-Potential-Dependent Gap Opening at the Dirac Surface States of Bi$_2$Se$_3$ Induced by Aggregated Substitutional Cr Atoms. {\it Phys. Rev. Lett.} {\bf 2014} {\it 112}, 056801.
 \bibitem{zhang} Zhang, G. et al. Quintuple-layer epitaxy of thin films of topological insulator Bi$_2$Se$_3$. {\it Appl. Phys. Lett.} {\bf 2009} {\it 95}, 053114.
 \bibitem{sakamoto} Sakamoto, Y., Hirahara, T., Miyazaki, H., Kimura, S.-I. \& Hasegawa, S. Spectroscopic evidence of a topological quantum phase transition in ultrathin Bi$_2$Se$_3$ films. {\it Phys. Rev. B} {\bf 2010} {\it 81}, 165432.
 \bibitem{xue} Zhang, Y. et al. Crossover of the three-dimensional topological insulator Bi$_2$Se$_3$ to the two-dimensional limit. {\it Nature Phys.} {\bf 2010} {\it 6}, 584.
 \bibitem{samrai} Kimura, S.-I. et al. SAMRAI: A novel variably polarized angle-resolved photoemission beamline in the VUV region at UVSOR-II. {\it Rev. Sci. Instrum.} {\bf 2010} {\it 81}, 053104.
 \bibitem{okuda} Okuda, T. et al. Efficient spin resolved spectroscopy observation machine at Hiroshima Synchrotron Radiation Center. {\it Rev. Sci. Instrum.} {\bf 2011} {\it 82}, 103302.
 \bibitem{vanhove} Van Hove, M. A. et al. Automated determination of complex surface structures by LEED. {\it Surf. Sci. Rep.} {\bf 1993} {\it 19}, 191.
 \bibitem{nakagawa} Nakagawa, T.; Takagi, Y.; Matsumoto, Y; Yokoyama, T., Enhancements of Spin and Orbital Magnetic Moments of Submonolayer Co on Cu(001) Studied by X-ray Magnetic Circular Dichroism Using Superconducting Magnet and Liquid He Cryostat. {\it Jpn. J. Appl. Phys.} {\bf 2008} {\it 47}, 2132.
 \bibitem{kresse1} Kresse, G.; Hafner, J., {\it Ab initio} molecular dynamics for open-shell transition metals. {\it Phys. Rev. B} {\bf 1993} {\it 48}, 13115.
 \bibitem{kresse2} Kresse, G. \& Furthm\"{u}ller, J. Efficient iterative schemes for {\it ab initio} total-energy calculations using a plane-wave basis set. {\it Phys. Rev. B} {\bf 1996} {\it 54}, 11169.
 \bibitem{perdew} Perdew, J. P.; Burke, K.; Ernzerhof, M., Generalized gradient approximation made simple. {\it Phys. Rev. Lett.} {\bf 1996} {\it 77}, 3865.
 \bibitem{blochl}  Bl\"{o}chl, P. E., Projector augmented-wave method. {\it Phys. Rev. B} {\bf 1994} {\it 50}, 17953.
 \bibitem{kresse3} Kresse, G.; Joubert, D., From ultrasoft pseudopotentials to the projector augmented-wave method. {\it Phys. Rev. B} {\bf 1999} {\it 59}, 1758.
 \bibitem{koelling} Koelling, D. D.; Harmon, B. N., A technique for relativistic spin-polarised calculations. {\it J. Phys. C} {\bf 1977}{\it 10}, 3107.
 \bibitem{liechtenstein} Liechtenstein, A. I.; Anisimov, V. I.; Zaanen, J., Density-functional theory and strong interactions: Orbital ordering in Mott-Hubbard insulators. {\it Phys. Rev. B} {\bf 1995} {\it 52}, R5467.
 \bibitem{Grimme} Grimme, S.; Antony, J.; Ehrlich, S.; Krieg, H., A consistent and accurate {\it ab initio} parametrization of density functional dispersion correction (DFT-D) for the 94 elements H-Pu. {\it J. Chem. Phys.} {\bf 2010} {\it 132}, 154104.
 \bibitem{Soluyanov} Soluyanov, A. A.; Vanderbilt, D., Computing topological invariants without inversion symmetry. {\it Phys. Rev. B} {\bf 2011} {\it 83}, 235401.
}
\end{thebibliography}

\end{document}